\documentclass[twocolumn,printnumbers,amsmath,amssymb,prl]{revtex4}
\usepackage{graphicx}
\usepackage{color}

\begin{document}
\title{Instabilities of jammed packings of frictionless spheres under compression}

\date{\today}

\author{Ning Xu$^{1}$}
\author{Andrea J. Liu$^2$}
\author{Sidney R. Nagel$^3$}

\affiliation{$^1$CAS Key Laboratory of Soft Matter Chemistry, Hefei National Laboratory for Physical Sciences at the Microscale, and Department of Physics, University of Science and Technology of China, Hefei 230026, P. R. China; \\$^2$Department of Physics and Astronomy, University of Pennsylvania, Philadelphia, PA 19104, USA; \\$^3$Department of Physics and James Franck and Enrico Fermi Institutes, The University of Chicago, Chicago, Illinois 60637, USA}

\begin{abstract}
We consider the contribution to the density of vibrational states and the distribution of energy barrier heights of incipient instabilities in a glass modeled by a jammed packing of spheres. On approaching an instability, the frequency of a normal mode and the height of the energy barrier to cross into a new ground state both vanish. These instabilities produce a contribution to the density of vibrational states that scales as $\omega^3$ at low frequencies $\omega$, but which vanishes in the thermodynamic limit. In addition, they affect an anharmonic property, the distribution of energy barriers $\Delta H$, giving a contribution that scales as $\Delta H^{-1/3}$ at low barrier heights, which should be universal for jammed and glassy systems.
\end{abstract}

\maketitle

Disordered solids inhabit an extremely high-dimensional rugged energy landscape with a vast number of metastable minima~\cite{stillinger,debenedetti}.  Dealing with such a complex topography poses challenges for understanding how the system moves among metastable basins as a result of thermal energy or an external perturbation such as compression or shear.  Here we focus on instabilities induced in zero-temperature ($T=0$) systems by compression as a first step in understanding how a disordered solid traverses its landscape at very low temperatures.

When a system in a given energy basin becomes unstable at $T=0$, a potential-energy barrier must vanish along some direction of configurational space, and the basin must flatten out in this direction.  This vanishing curvature corresponds to the vanishing frequency of a vibrational mode \cite{malandro,maloney,networkremark}.  As a result, low-frequency vibrational modes have been shown to contain predictive information about incipient instabilities~\cite{manning} and such instabilities are associated with low-frequency quasilocalized modes with particularly low energy barriers~\cite{ning,ning1,vincenzo,degiuli}.  Zero-temperature jammed sphere packings provide special insight into this physics because they are nearly marginally stable~\cite{arcmp,matthieuthesis,matthieureview}; at every pressure $p$, the system is nearly unstable to compression, implying that there is an instability at a nearby higher pressure.    At low temperatures, one would expect the lowest energy barriers to control how the system explores its landscape, so the zero-temperature study of instabilities provides a way of gaining insight into low-temperature behavior beyond the harmonic approximation.

In this paper, we calculate for jammed packings the distribution of mode frequencies $\omega$ and energy barriers $\Delta H$ corresponding to the most vulnerable directions in the energy landscape for encountering an instability. These distribution functions vary as a function of the distance to the jamming transition.   We find that the contribution of incipient instabilities to the vibrational spectrum scales as $\omega^3$. However, we also find that the number of compression instabilities per strain scales sublinearly with system size $N$, leading to a contribution to the density of vibrational states that vanishes in the thermodynamic limit.   However, results by Salerno and Robbins~\cite{salerno1} on shear instabilities, combined with our calculations, suggest that shear instabilities could give rise to a contribution that survives in the thermodynamic limit and scales as $\omega^3$ . We also find that the distribution of energy barriers corresponding to incipient instabilities obeys a power law, $P_H(\Delta H) \propto \Delta H^{-1/3}$. Thus, the landscape appears to have a fractal character in three dimensions.
\begin{figure}
\vspace{0.05in}
\includegraphics[width=0.48\textwidth]{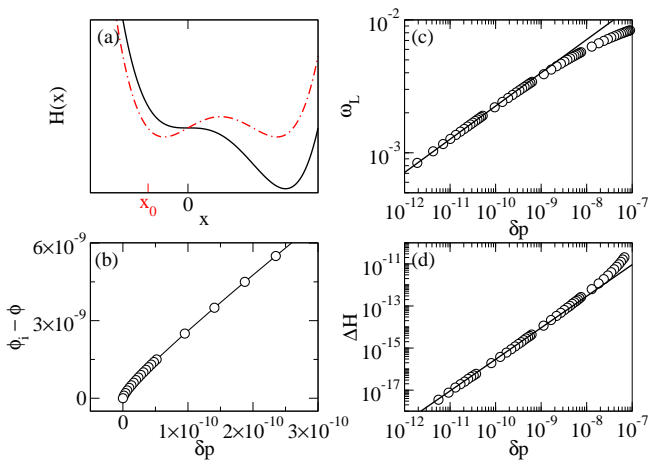}
\caption{\label{fig:fold} (a) Schematic plot of the enthalpy landscape $H$ along the reaction coordinate $x$ at the pressure corresponding to the instability, $p_i$ (solid), and at a pressure below the instability, $p<p_i$ (dot-dashed), where the minimum is shifted to $x_0$. (b) An example of the response of packing fraction $\phi$ to the increase of pressure $p$ upon approaching the instability at $\phi_i$ and $p_i$ from below, where $\delta p \equiv p_i-p$ is the distance to the instability. The line is a guide for the eye. (c)-(d) Frequency $\omega_L$ and enthalpy barrier height $\Delta H$ for the mode associated with the instability as a function of $\delta p$. The lines in (c) and (d) indicate the expected power law scalings $\omega_L \sim \delta p^{1/4}$ and $\Delta H \sim \delta p^{3/2}$, respectively.}
\end{figure}

{\it Fold instability:}  We start by reviewing the expected scaling of the lowest vibrational mode frequency and barrier height as the system is compressed towards the pressure $p_i$ at which the instability occurs.  Because we control the pressure, the relevant landscape is the enthalpy landscape.  Consider the enthalpy of a system, $H$, along the ``reaction coordinate" $x$ in configurational space.  Assume that the saddle point is at $x=0$ when $p=p_i$, as shown in Fig.~\ref{fig:fold}(a). There is no term linear in $x$.  Moreover, since the system is unstable at $p_i$, the curvature (or the square of the mode frequency) must vanish, so there is also no quadratic term in $x$.  $H$ must therefore generically be cubic in $x$. We now step back from the instability, so that the system is at a pressure $p<p_i$.  The lowest non-vanishing coupling term between $\delta p \equiv p_i-p$ and $x$ is generically linear, so
\begin{equation}
H= -\frac{1}{3} a_3 x^3 + c x \delta p .\label{eq:U1}
\end{equation}
Here we assume that $a_3>0$, so that at $\delta p=0$ (at the instability) the system is unstable towards a minimum that lies at $x>0$.   The linear term shifts the minimum to $x_0 = -\sqrt{c {\delta p}/a_3}$ ($c>0$), as shown in Fig.~\ref{fig:fold}(a).  Expanding $H$ around $x=x_0$ in Eq.~(\ref{eq:U1}), we find
\begin{equation}
H=\frac{2}{3}a_3x_0^3-a_3 x_0 (x-x_0)^2 - \frac{1}{3}a_3 (x-x_0)^3 + {\cal O}[(x-x_0)^4].
\end{equation}
Therefore $\omega_L^2 = -2a_3x_0=2\sqrt{c a_3{\delta p}}$, where $\omega_L$ is the frequency of the mode associated with the instability.  Thus,
\begin{equation}
\omega_L = (4c a_3)^{1/4} (\delta p)^{1/4}. \label{eq:omegaL}
\end{equation}
This typical scaling of a fold instability~\cite{foldinstability} has long been observed in studies of soft modes in glassy systems \cite{malandro,maloney}.
The energy-barrier height scales as
\begin{equation}
\Delta H = \frac{4}{3} \frac{c^{3/2}}{a_3^{1/2}} \delta p^{3/2}. \label{eq:DeltaH}
\end{equation}

{\it Simulations:}  We confirm this scaling using numerical simulations of three-dimensional systems of $N\in[250,4000]$ frictionless spheres.
Specifically, we consider jammed packings with a Hertzian interaction potential between particles $i$ and $j$:
\begin{equation}
V(r_{ij})=\frac{2\epsilon}{5}\left( 1-\frac{r_{ij}}{\sigma_{ij}}\right)^{5/2}\Theta\left(1-\frac{r_{ij}}{\sigma_{ij}}\right), \label{eq:hertzian}
\end{equation}
where $r_{ij}$ and $\sigma_{ij}$ are the separation between particles $i$ and $j$ and sum of their radii respectively, $\epsilon$ is the characteristic energy, and $\Theta(x)$ is the Heaviside step function.  (We use the Hertzian  instead of the harmonic potential so that there is no discontinuity in the second derivative of the potential at the point of contact.)  Periodic boundary conditions are applied in all directions.  We study a 50:50 mixture of particles with diameters $\sigma$ and $\sigma_L=1.4\sigma$, respectively. All particles have the same mass $m$. The units of length, mass, and energy are $\sigma$, $m$, and $\epsilon$.  The frequency is in the units of $\sqrt{\epsilon/m\sigma^2}$.

By rapidly quenching ideal-gas states to zero temperature using a fast inertial relaxation engine algorithm~\cite{fire}, we minimize the enthalpy at fixed pressure to obtain mechanically stable disordered solids. For each system size, we generate 5000 distinct states at the desired pressure $p$. We then compress each state by increasing $p$ while minimizing the enthalpy $H$, until there is an abrupt change of packing fraction corresponding to the first instability for that initial state. To calculate the vibration modes, we diagonalize the Hessian matrix corresponding to the enthalpy using ARPACK~\cite{arpack}.

Figures~\ref{fig:fold}(b)-(d) show the approach to an instability at $p_i$.  The packing fraction $\phi$ is shown versus pressure in Fig.~\ref{fig:fold}(b).  Note that at $p=p_i$ ($\delta p = 0$) the bulk modulus $B=\phi \frac{{\rm d}P}{{\rm d}\phi} =0$ \cite{note,dagois-bohy}.

The frequency of the mode associated with the instability, $\omega_L$, is shown in Fig.~\ref{fig:fold}(c) as we decrease $\delta p$ by raising the pressure $p$ towards $p_i$.  As expected, $\omega_L \sim \delta p^{1/4}$.  The same scaling is obtained for the Lennard-Jones glass~\cite{maloney}. In addition, Fig.~\ref{fig:fold}(d) shows that the height of the enthalpy barrier vanishes as the instability is approached as $\Delta H \sim \delta p^{3/2}$ as expected.

\begin{figure}
\vspace{-0.in}
\center
\includegraphics[width=0.48\textwidth]{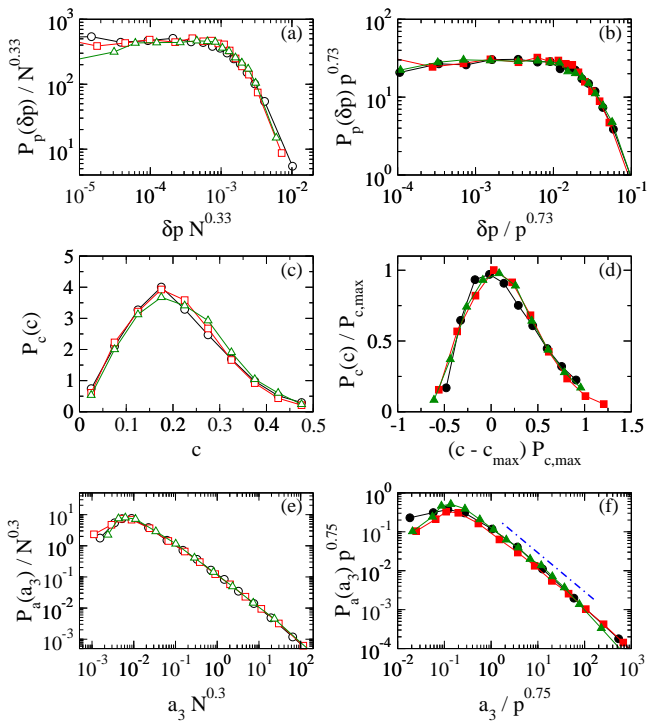}
\caption{\label{fig:distbn} (Left column) System-size and (Right column) pressure dependence of the distributions of $\delta p$, $c$ and $a_3$ characterizing the instabilities.  For the left column, system sizes are $N=250$ (circles), $1000$ (squares) and $4000$ (triangles), all at $p=10^{-3}$. For the right column, pressures are $p=10^{-4}$ (circles), $10^{-3}$ (squares) and $10^{-2}$ (triangles), all at $N=1000$.  Solid lines are guides for the eye. The dot-dash line in (f) has a slope of $-1$. In (d), peak position $c_{\textrm max}$ and amplitude $P_{\textrm{c,max}}$ are $(0.0989,5.15)$, $(0.163,3.92)$ and $(0.197,3.49)$ at pressures $p=10^{-4}, 10^{-3}$ and $10^{-2}$, respectively.}
\end{figure}

{\it Distribution functions:}  The values of the distance to the nearest instability, $\delta p$ and the variables $c$ and $a_3$ defined in Eq.~(\ref{eq:U1}), vary from one enthalpy minimum to another.  We characterize the ensemble of instabilities in terms of the distributions of the distance to the nearest instability, $P_{p}(\delta p)$, the coupling constant $c$, $P_c(c)$, and of the cubic coefficient $a_3$, $P_a(a_3)$, for an ensemble of initial enthalpy minima for three-dimensional packings.  Figure~\ref{fig:distbn}(a) shows that $P_p(\delta p)$ is approximately flat with increasing $\delta p$ until it falls off at $\delta p_y$.  Because we are measuring only the distance from $p$ to the \emph{first} instability (at $p_i$), $\delta p_y$ corresponds to the yield strain for compression and is comparable to the pressure interval between instabilities. The distributions for different system sizes collapse when $\delta p$ is scaled by $N^{0.33\pm 0.05}$ [Fig.~\ref{fig:distbn}(a)].  This implies that $\delta p_y$ decreases as $N^{-0.33 \pm 0.05}$ with increasing $N$.  Sublinear scaling in $1/N$ of the distance between instabilities has been observed in several contexts~\cite{maloney1,karmakar,salerno,salerno1} and is associated with an exponent $\theta>0$ characterizing the distribution of yield stresses as well as the existence of avalanches~\cite{lin}, such that $\delta p_y \sim N^{-1/(\theta+1)}$. Sublinear scalings in $1/N$ are also seen in calculations of energy barriers in mean-field spin glasses~\cite{parisi,moore}.  In spin glasses, this decrease may be associated with the fractal nature of the energy landscape  -- a feature also predicted for jammed packings~\cite{charbonneau}.

Fig.~\ref{fig:distbn}(b) shows how $P_{p}(\delta p)$ depends on the initial pressure $p$: $\delta p_y \sim p^{0.73 \pm 0.03}$.  As expected, the compressive yield strain vanishes as the jamming transition is approached.

Figure~\ref{fig:distbn}(c) shows that the distribution of $c$ is independent of $N$ and is fairly innocuous with a peak at $c \approx 0.16$ when $p=10^{-3}$. Figure~\ref{fig:distbn}(d) shows that the shape of the $P_c(c)$ distribution does not change appreciably with pressure.

The distribution of cubic coefficients, $P_a(a_3)$ is shown in Figs.~\ref{fig:distbn}(e) and \ref{fig:distbn}(f). There is a broad, approximately power-law, distribution of $a_3$ that is cut off at low values of $a_3$.  As $N$ increases, the minimum shifts to lower $a_3$ as $N^{-0.3 \pm 0.1}$.  As the pressure is reduced towards the jamming transition, the distribution appears to approach a pure power law of $P(a_3) \sim a_3^{-1}$ at small $a_3$, as the peak value vanishes approximately as $p^{0.75 \pm 0.05}$.

{\it Density of states and energy barriers:}  We now consider the contribution of instabilities to the density of vibrational states and the distribution of energy barriers at a given pressure $p$.  An incipient instability at $p_i>p$ will contribute a mode at a frequency given by Eq.~(\ref{eq:omegaL}) and a barrier height given by Eq.~(\ref{eq:DeltaH}). The contribution of instabilities to the density of states or to the energy-barrier distribution is therefore the sum over the contributions of all of the incipient instabilities. The number of instabilities in a given pressure interval is ${\cal P}\propto 1/\delta p_y \sim N^{0.33} p^{-0.73}$.  The contribution of instabilities to the normalized density of states is then
\begin{eqnarray}
 D(\omega)&=&\frac{{\cal P}}{3N} \int_p^\infty dp_i \int_0^\infty da_3 P_a(a_3) \nonumber\\
& &\int_0^\infty dc P_c(c)\delta (\omega-\sqrt{2} (c a_3 \delta p)^{1/4}),
\end{eqnarray}
and the contribution to the distribution of energy barriers is
\begin{eqnarray}
P_H(\Delta H)=& &{\cal P} \int_p^\infty dp_i \int_0^\infty da_3 P_a(a_3) \int_0^\infty dc P_c(c) \nonumber \\& &\delta(\Delta H-\frac{4}{3} c^{3/2} a_3^{-1/2} (\delta p)^{3/2}).
\end{eqnarray}
In the low-frequency limit, this leads to
\begin{equation}
D(\omega)= \frac{{\cal P} \langle a_3^{-1} \rangle \langle c^{-1} \rangle}{3N} \omega^3, \label{eq:dosfinal}
\end{equation}
where $\langle a_3^{-1} \rangle = \int_0^\infty P_a(a_3) a_3^{-1}{\rm d}a_3$ and $\langle c^{-1} \rangle = \int_0^\infty P_c(c) c^{-1}{\rm d}c$.  Equation~(\ref{eq:dosfinal}) shows that instabilities give rise to an $\omega^3$ 
contribution to the density of states. From the scalings in Fig.~\ref{fig:distbn}, we see that the contribution to the density of states vanishes as $N^{-0.37 \pm 0.15}p^{-1.48 \pm 0.08}$ as $N \rightarrow \infty$. 
Thus, compression instabilities do not contribute to the density of normal modes of vibration in the thermodynamic limit. 

For low energy barriers, we find
\begin{equation}
P_H(\Delta H)= (\frac{1}{6})^{1/3} {\cal P}\langle a_3^{1/3}\rangle \langle c^{-1}\rangle  \Delta H^{-1/3} .
\end{equation}
Note that if $P_a(a_3) \sim 1/a_3$ at large $a_3$, as suggested by the dot-dash line in Fig.~\ref{fig:distbn}(f), then $\langle a_3^{1/3} \rangle$ diverges at the high $a_3$ (low energy barrier) end.  A closer look at Fig.~\ref{fig:distbn}(f), however, suggests that $P_a(a_3)$ bends down more rapidly than $1/a_3$ at high $a_3$. We therefore assume that $\langle a_3^{1/3} \rangle$ is finite.  In that case, $P_H$ increases with ${\cal P} \langle a_3^{1/3} \rangle \sim N^{0.23 \pm 0.08} p^{-0.48 \pm 0.05}$. Like the density of states, this distribution also diverges as $p \rightarrow 0$ signaling a breakdown of the fold instability description.

{\it Discussion and conclusions:}  The main assumptions of our analysis have been that a disordered solid is generically always close to a compressive instability -- that is, it is nearly marginally stable with respect to compression.  The issue of marginal stability has been studied in systems consisting of jammed packings of spheres.  A packing of spheres with soft, finite-ranged repulsions is marginally stable at the jamming transition at zero pressure, where there are the minimum number of contacts needed for mechanical stability~\cite{arcmp}.  As the system is compressed above the jamming transition, the connectivity increases, but at a pressure $p$, the system is still close to the limit of stability with respect to compression -- it is just stable enough to support that pressure~\cite{matthieuthesis,matthieureview}.

Note that similar instabilities arise because the system is nearly marginally stable to shear stress, and that a full accounting for the contribution of instabilities due to all types of loading conditions would lead to larger contributions to the density of states and distribution of energy barriers. Potentially these would have different sublinear scalings with system size.  In particular, it has been shown~\cite{salerno1} that the typical strain separating instabilities in Lennard-Jones systems scales as $N^{-0.62 \pm 0.08}$, implying that ${\cal P} \sim N^{0.62 \pm 0.08}$.  From Eq.~(\ref{eq:dosfinal}) we see that for shear instabilities, the $N$-dependence of the contribution to the density of states scales as $N^{-0.1 \pm 0.2}$ so that there is no $N$-dependence to within measurement error. This suggests that shear instabilities may give a contribution to the density of states that scales as $\omega^3$ that survives in the thermodynamic limit.

This contribution would be smaller at low frequencies than the Debye contribution, which scales as $\omega^2$. A contribution of $\omega^3$ has not been observed directly in the three-dimensional density of states~\cite{epitome,leo}; Charbonneau, et al.~\cite{corwin2016} report a low-frequency density of states that scales as $\omega^2$, consistent with mean-field expectations~\cite{franz}, while Lerner, et al.~\cite{lerner2016} find $\omega^4$.  Our result suggests that there may be a regime of frequencies, above the Debye-dominated regime but below currently accessible frequencies, where the density of states scales as $\omega^3$.  It has been demonstrated in simulations that low-frequency, localized, anharmonic modes in jammed systems can produce echo phenomena~\cite{Burton}. If this contribution to the density of states does persist in the thermodynamic limit, then these modes might provide a source for the phonon echoes observed in experiments but which have previously been ascribed to quantum-mechanical two-level systems~\cite{Golding}.

Near the jamming transition, marginal stability with respect to connectivity leads to instabilities that are extended, not localized, and gives rise to a plateau in the vibrational spectrum coming from extended, disordered vibrational modes, known as the boson peak, at frequency $\omega^*$~\cite{matthieuthesis}.  In that limit, we find that the contributions of instabilities to the density of states and to the energy barrier distribution diverge because ${\cal P} \langle a_3^{-1} \rangle$ and ${\cal P} \langle a_3^{1/3} \rangle$ diverge.  This could be a sign that our picture based on cubic instabilities breaks down near the jamming transition. We have also found that a number of anharmonic properties, including the compressive yield strain and the distribution of cubic coefficients, appear to obey power-law scaling with pressure near the jamming transition, with new exponents.

We have shown that cubic instabilities lead to an $\omega^3$ contribution to the low-frequency density of states and a $\Delta H^{-1/3}$ contribution to the distribution of low energy barriers. In mean-field, instabilities lead to an $\omega^2$ contribution to the low-frequency density of states \cite{franz}. It is therefore unlikely that the power law that we have identified for the energy barrier distribution is the same as the one predicted by mean-field calculations~\cite{charbonneau}.

While our results have been derived in the context of jammed packings, they are likely to be more generally valid.  Near-marginal stability is not specific to packings of jammed spheres but should be a generic feature of glassy materials.   The distribution of cubic coefficients, $P(a_3)$, is likely to be robust, since earlier calculations~\cite{ning} suggest that systems such as Lennard-Jones glasses are not far from marginal stability, but this also needs to be tested in order to understand the implications of our results for real materials.  The predicted power law scalings for the energy barrier distribution and density of states originate from the scaling of the fold instability, which is universal. One remarkable conclusion from these results is that the nature of the ground states in jammed systems has universal anharmonic as well as harmonic properties. The existence of universal anharmonic features has previously been hinted at in studies of the energy barriers in the directions of the normal modes in jammed systems~\cite{ning1}.

We thank S. S. Schoenholz, D. M. Sussman, J. P. Sethna, F. Zamponi and G. Parisi for instructive discussions.  This work was supported by the National Natural Science Foundation of China Grants No.~21325418 and No.~11574278 (NX), Fundamental Research Funds for the Central Universities Grant No.~2030020028 (NX), and by the Simons Foundation for the collaboration "Cracking the Glass Problem" (454945 to AJL and 348125 to SRN) and the U.S. Department of Energy, Office of Basic Energy Sciences, Division of Materials Sciences and Engineering under Grants No.~DE-FG02-05ER46199 (AJL, NX) and No.~DE-FG02-03ER46088 (SRN,NX).


\begin{thebibliography}{50}
\bibitem{networkremark} Note that these instabilities are distinct from events that change the contact network but do not lead to the vanishing of a low-frequency mode~\cite{manning}.
\bibitem{note} In earlier studies of sheared glasses~\cite{maloney}, in which instabilities were induced by applying shear, the shear modulus actually became negative before the instability. This unphysical result arises because the system was held at fixed volume and box shape~\cite{dagois-bohy}.
\end{thebibliography}
\end{document}